\documentclass[a4paper,12pt]{article}
\usepackage{jheppub}
\usepackage{lineno}
\usepackage{microtype}

\newcommand\td{\text{d}}
\newcommand\tD{\text{D}}
\newcommand{\p}{\partial}
\newcommand{\bo}{\boldsymbol}

\def\>{\rangle} \def\<{\langle}

\title{\boldmath Symmetries and Critical Dimensions of Tensionless Branes}

\author[a,b,c]{Bin Chen}
\author[b]{Zezhou Hu}

\affiliation[a]{Institute of Fundamental Physics and Quantum Technology, \\\&  School of Physical Science and Technology, \\ Ningbo University, Ningbo, Zhejiang 315211, China}
\affiliation[b]{School of Physics, Peking University, \\No.5 Yiheyuan Rd, Beijing 100871, China}
\affiliation[c]{Center for High Energy Physics, Peking University, \\No.5 Yiheyuan Rd, Beijing 100871, China}

\emailAdd{chenbin1@nbu.edu.cn, z.z.hu@pku.edu.cn}

\abstract{
In this work, we investigate the worldsheet symmetry of bosonic brane theories and its quantum consistency in the tensionless limit. We find that the residual worldsheet symmetry after specific gauge fixing is generated by a novel algebra, denoted as $g^{(p)}_\lambda$. To achieve full quantization of the tensionless brane, we introduce a $bc$ ghost system and derive the overall BRST charge. Moreover, we calculate the quantum anomaly of the $g^{(p)}_\lambda$ algebra for general parameters $p$ and $\lambda$ in the framework of canonical quantization. After demanding that this quantum anomaly vanishes, we derive the critical dimensions of the bosonic brane theories. Especially, we obtain nontrivial solutions for the tensionless string: $D=14$ spacetime dimensions when $\lambda=-1$ and $D=26$ spacetime dimensions when $\lambda=1$. We also notice that the quantum anomaly automatically vanishes after the Riemann zeta function regularization for $p>1$. This is consistent with the mathematical fact that there is no scalar central extension of the algebra $g^{(p)}_\lambda$ for $p>1$.


}

\begin{document}
\maketitle
\flushbottom

\section{Introduction}\label{sec:intro}

For a theory of $p$-brane(string) living in a flat background with Minkowski metric $\eta_{\mu\nu}$, there is Nambu-Goto action
\begin{equation}\label{eq:NG}
    S=T_p\int\td^d\sigma \sqrt{|G_{\alpha\beta}|}\,,
\end{equation}
where $G_{\alpha\beta}$ is the induced metric\footnote{In this paper, the Greek letters $\mu,\nu$ varying from 0 to $D-1$ with $D$ being the spacetime dimension. And the Greek letters $\alpha,\beta,\gamma,\cdots$ vary from $0$ to $p$, and the Latin letters $i,j,k,\cdots$ vary from $1$ to $p$. The total dimension of the $p$-brane worldsheet is $d=p+1$. In the main body of this paper, we sometimes express the spacetime vector as $\bo{X}\equiv X^\mu \p_\mu$ and the worldsheet vector as $\hat{\epsilon}\equiv \epsilon^\alpha\p_\alpha$ for simplicity.}
\begin{equation}
    G_{\alpha\beta}=\eta_{\mu\nu}\p_\alpha X^\mu\p_\beta X^\nu\,,
\end{equation}
and the worldsheet coordinates are given by $\sigma^\alpha=(t,\vec\sigma)$ and $\vec\sigma=\{\sigma^i|i=1,2,\cdots,p\}$. Furthermore, this action is classically equivalent to the Polyakov action
\begin{equation}\label{eq:Polyakov}
    S=-\frac{T_p}{2}\int \td^d \sigma \sqrt{-h}h^{\alpha\beta}\p_\alpha X\cdot\p_\beta X+\frac{(p-1)T_p}{2}\int\td^d\sigma\sqrt{-h}\,,
\end{equation}
with $h_{\alpha\beta}$ being the auxiliary metric on the worldsheet. It is usually easier to quantize the theory from the Polyakov action.
In the $p=1$ case of string theory \cite{Green:2012oqa}, the metric $h_{\alpha\beta}$ has three independent components. None of them are dynamical variables because of the two-dimensional worldsheet diffeomorphism and the one-dimensional Weyl invariance of the action, which renders the equations of motion (E.o.M.) of string theory fully linear. And after classical covariant gauge fixing, the residual worldsheet symmetry is generated by the Virasoro algebra. Then we can obtain the critical dimension by requiring the quantum anomaly of the Virasoro symmetry to vanish. However, such a strategy does not work for $p>1$ brane theory due to the intrinsic non-linearity. There, the auxiliary metric is a symmetric $(p+1)\times(p+1)$ tensor, and there is only a $(p+1)$-dimensional diffeomorphism such that  $h_{\alpha\beta}$ has nontrivial dynamics, and the total system is highly non-linear.

There has been long-standing interest in studying $p>1$ brane theory. In mid-1980, by including a Wess-Zumino-Witten term in membrane action, it has been shown that  supermembranes \cite{Hughes:1986fa, Achucarro:1987nc} should live in the 11-dimensional background\cite{Duff:1987bx,Bergshoeff:1987cm, Bergshoeff:1987dh, Bars:1987dy}. Still, it is very difficult to deal directly with the quantum consistency of worldsheet diffeomorphism and fermionic Siegel symmetry, and to the best one can only work under the light-cone gauge \cite{Hoppe1982}.  The efforts trying to perform quantization of the supermembrane include \cite{Duff:1987cs, Fujikawa:1987av, Bars:1988uj}. Later on, researchers turned to explore the non-perturbative aspects of supermembranes by using matrix models \cite{deWit:1988wri} and found the supermembrane is unstable \cite{deWit:1988xki}. In 1990s, the discovery of mysterious M-theory\cite{Witten:1995ex} rekindled the interests in membrane\cite{Townsend:1995kk}, matrix model\cite{Banks:1996vh}, and M5-brane\cite{Aganagic:1997zq}.

In this work, we would like to investigate the higher-dimensional brane in the tensionless limit, and we start from the bosonic case. In the tensionless limit, the theory gets linearized: the second term in \eqref{eq:Polyakov} is effectively vanishing, and the first term can be well described by an ILST-type action \cite{Isberg:1993av}, which is linear.  For $p=1$ case, the tensionless (super)string \cite{Bagchi:2016yyf, Bagchi:2017cte, Bagchi:2020fpr, Chen:2023esw, Bagchi:2024qsb} has already been explored for some years, see \cite{Bagchi:2026wcu} for a nice review. It has been shown \cite{Demulder:2023bux} that the residual worldsheet symmetry of the tensile string gets modified from Virasoro algebra to $\mathfrak{bms}_3$ (or $\mathfrak{g}^{(1)}_1$ in our language). And its T-duality partner, Carrollian string are also studied broadly \cite{Gomis:2023eav, Bagchi:2024rje, Chen:2025gaz, Figueroa-OFarrill:2025njv}. For higher dimensional cases, the quantization of the null brane was studied in \cite{Saltsidis:1997nx, Bozhilov:1997xq} and was recently explored in the lightcone gauge by \cite{Dutta:2024gkc}. Another recent work \cite{Borsten:2025dcx} studying brane models demonstrates an enhancement of symmetry algebras in the tensionless limit. Our study will focus on the quantum consistency of the worldvolume symmetry.

The remaining parts of this note are organized as follows. In Section \ref{sec:symmetry}, we show that after some gauge fixing, the residual worldvolume symmetry is generated by  the algebra  $\mathfrak{g}^{(p)}_{\lambda}\cong \text{Vect}(T^p)\ltimes_\lambda C^\infty(T^p)$, where $\lambda$ is a free parameter and different $\lambda$s correspond to different gauge transformations. Furthermore we construct the generators of the algebra $\mathfrak{g}^{(p)}_{\lambda}$ in terms of the modes of matter fields in the formalism of canonical quantization.
In Section \ref{sec:bcghost}, we consider path-integral quantization of the tensionless brane, and thereafter introduce $b c$ ghost system. We obtain the BRST charge of the ghost system and the $\mathfrak{g}^{(p)}_{\lambda}$ generators for the ghost fields. In Section \ref{sec:QuantumAnomaly}, we calculate the quantum anomaly of the algebra $\mathfrak{g}^{(p)}_{\lambda}$ for general $p$ and $\lambda$. After all, we obtain the critical dimensions of the tensionless branes in Section \ref{sec:conclusions}. We also give some comments on the conclusions and the future directions of the novel algebra $\mathfrak{g}^{(p)}_{\lambda}$.

\section{The worldsheet symmetry of the tensionless brane: \texorpdfstring{$\mathfrak{g}^{(p)}_{\lambda}$}{g(p)lambda}}\label{sec:symmetry}

The direct tensionless limit of the Nambu-Goto action \eqref{eq:NG} is ill-defined. The action of the tensionless brane(string) is actually given by the ILST action\cite{Isberg:1993av}, shown as
\begin{equation}
    S=\left(\frac{1}{2\pi}\right)^p\int \td^{d}\sigma \frac{1}{2}V^{\alpha}V^\beta \p_\alpha \bo{X}\cdot\p_\beta \bo{X},\quad p\geq1\,.
\end{equation}
The action has a reparameterization invariance (diffeomorphism) given by
\begin{equation}
    \begin{aligned}
        &\delta_{\hat{\epsilon}} \bo{X}=-\epsilon^\alpha\p_\alpha \bo{X}\,,\\
        &\delta_{\hat{\epsilon}} V^\alpha=-\epsilon^\beta\p_\beta V^\alpha+V^\beta\p_\beta \epsilon^\alpha-\frac{1}{2}\p_\beta\epsilon^\beta V^\alpha\,,
    \end{aligned}
\end{equation}
and a local Weyl-like invariance given by
\begin{equation}
    \begin{aligned}
        &\delta_{\chi} \bo{X}=-\frac{\Delta}{d}\chi \bo{X}\,,\\
        &\delta_{\chi} V^\alpha=+\frac{\Delta}{d}\chi V^\alpha\,,
    \end{aligned}
\end{equation}
with $\chi=\chi(\sigma)$ satisfying $V^\beta\p_\beta\chi=0$. If the diffeomorphism $\hat{\epsilon}$ also satisfy
\begin{equation}\label{eq:CarrDiff}
    V^\beta\p_\beta(\p_\alpha\epsilon^\alpha)=0\,,
\end{equation}
then we can linearly combine the two local invariances above by setting $\chi=\p_\beta\epsilon^\beta$. Under such a combined local transformation, the fields transform as
\begin{equation}
    \begin{aligned}
        &\delta_{\hat{\epsilon}} \bo{X}=-\epsilon^\alpha\p_\alpha \bo{X}-\frac{\Delta}{d}\p_\alpha\epsilon^\alpha \bo{X}\,,\\
        &\delta_{\hat{\epsilon}} V^\alpha=-\epsilon^\beta\p_\beta V^\alpha+V^\beta\p_\beta \epsilon^\alpha-\left(\frac{1}{2}-\frac{\Delta}{d}\right)\p_\beta\epsilon^\beta V^\alpha\,,
    \end{aligned}
\end{equation}
where the parameter $\Delta$ labeling the conformal dimension of the matter field $X$ can be chosen arbitrarily due to the condition \eqref{eq:CarrDiff}.

The vielbein $V^\alpha$ serves as the geometric structure of the tensionless worldsheet. We can choose a gauge $\hat{V}=\hat{V}_{(\hat{0})}\equiv(1,\vec0)$. Under this gauge, we find a solution (but not the most general solution) to the condition \eqref{eq:CarrDiff},
\begin{equation}\label{eq:CarrDiffSub}
    \hat{\epsilon}=f^i(\vec\sigma)\p_i+\left(h(\vec\sigma)t+g(\vec\sigma)\right)\p_0\,,
\end{equation}
which is a subset of the Carrollian diffeomorphism\footnote{The Carrollian diffeomorphism is defined in \cite{Ciambelli:2018xat, Ciambelli:2019lap} as $t\to t'(t,\vec x)$ and $\vec x\to \vec x'(\vec x)$. And the associated infinitesimal diffeomorphism $x^\alpha\to x^\alpha+\epsilon^\alpha(x)$ is $\hat{\epsilon}=f^i(\vec\sigma)\p_i+g(t,\vec\sigma)\p_0$. Under such a diffeomorphism, the Carrollian structure $\{\p_0,\,g_{i j}\td x^i\td x^j\}$ remains intact.}.
The diffeomorphism that preserves the gauge $\hat{V}_{(\hat{0})}\equiv(1,\vec0)$ is the associated global symmetry, which is
\begin{equation}\label{eq:defAlg}
    \delta_{\hat{\xi}} V^\alpha|_{\hat{V}=\hat{V}_{(\hat{0})}}=0\,.
\end{equation}
Its general solution is
\begin{equation}
        \hat{\xi}=f^i(\vec\sigma)\p_i+\left(\lambda\p_k f^k(\vec\sigma) t+g(\vec\sigma)\right)\p_0\,,\quad \lambda=\frac{\frac{1}{2}-\frac{\Delta}{d}}{\frac{1}{2}+\frac{\Delta}{d}}\,.
\end{equation}
It is a special case of \eqref{eq:CarrDiffSub} with $h(\vec \sigma)$ set to $\lambda\p_k f^k(\vec\sigma)$.
For simplicity, we consider that the spatial manifold of the tensionless worldsheet is a torus $T^p$. Then the function $f^i(\vec\sigma),g(\vec\sigma)$ can be expanded by modes $e^{-i m_k \sigma^k}$. Finally, we can rewrite the global symmetry denoted by $\hat{\xi}$ as $-i \hat{l}^i_{\{m\}}$ and $-i \hat{m}_{\{m\}}$,
\begin{equation}\label{eq:CarrConVec}
    \begin{aligned}
        &\hat{l}^i_{\{m\}}=-i e^{-i m_k \sigma^k}\p_i-\lambda m_i t e^{-i m_k \sigma^k}\p_0\,,\\
        &\hat{m}_{\{m\}}=-i e^{-i m_k \sigma^k}\p_0\,,
    \end{aligned}
\end{equation}
which forms the $\mathfrak{g}^{(p)}_{\lambda}$ algebra
\begin{equation}
    \begin{aligned}
        &[\hat{l}^i_{\{m\}},\hat{l}^j_{\{n\}}]=m_j \hat{l}^i_{\{m+n\}}-n_i \hat{l}^j_{\{m+n\}}\,,\\
        &[\hat{l}^i_{\{m\}},\hat{m}_{\{n\}}]=(\lambda m_i-n_i) \hat{m}_{\{m+n\}}\,,
    \end{aligned}
\end{equation}
where the set $\{m\}$ is a short notation of $\{m_1,m_2,\cdots,m_p\}$ and $\{m+n\}$ is defined to be
$\{m_1+n_1,m_2+n_2,\cdots,m_p+n_p\}$. For $\lambda=1$, this algebra $g^{(p)}_{\lambda=1}$ and its possible central extension have been discussed in \cite{ Saltsidis:1997nx, Bozhilov:1997xq}. For $p=1,\lambda=1$, it is just the $\mathfrak{bms}_3$ algebra that appears in the usual tensionless string theory. In fact, the parameter $\lambda$ is responsible for the \textit{anisotropy} scaling symmetry in the Electric Carrollian free scalar. The \textit{isotropic} scaling symmetry is the special case $\lambda=\frac{1}{p}$.
For $p=1$, the algebra $\mathfrak{g}^{(1)}_{\lambda}$ with general $\lambda$ has already been discussed in \cite{Figueroa-OFarrill:2024wgs} with slightly different notations. For $p=1$ and $p=2$, the similar algebras involving $s$-supertranslation in \cite{Grumiller:2019fmp} could be related to $\mathfrak{g}^{(p)}_{\lambda}$ by $\lambda\sim\frac{s}{p}$.
For $p>1$ and $\lambda=\frac{1}{p}$, one can consider the subalgebra within \eqref{eq:CCAwithQA},
\begin{equation}
    \begin{aligned}
        &[L^i_m,L^i_n]=(m-n)L^i_{m+n}\,,\\
        &[L^i_m,L^j_n]=0\,,\quad (i\neq j)\,,\\
        &[L^i_m,M_n]=\left(\frac{m}{p}-n\right)M_{m+n}\,,
    \end{aligned}
\end{equation}
where
\begin{equation}\label{eq:BMSdef}
    L^i_m\equiv L^i_{\{m_i=m,m_{j\neq i}=0\}}\,.
\end{equation}
This is an extension of BMS$_{p+2}$ symmetry.

Next, we perform canonical quantization of the fields $\bo{X}$ under the gauge $\hat{V}_{(\hat{0})}\equiv(1,\vec0)$. Solving the equations of motion, we have the mode expansions
\begin{equation}
    \bo{X}=\sum_{\{m\}}(i \bo{A}_{\{m\}}+\bo{B}_{\{m\}} t)e^{i m_k \sigma^k}\,.
\end{equation}
Its canonical momentum is
\begin{equation}
    \bo{\Pi}=\left(\frac{1}{2\pi}\right)^p\p_t\bo{X}=\left(\frac{1}{2\pi}\right)^p\sum_{\{m\}}\bo{B}_{\{m\}}e^{i m_k \sigma^k}\,.
\end{equation}
From  the canonical commutation relation
\begin{equation}
    \begin{aligned}
        &[X^\mu(t,\vec\sigma),\Pi^\nu(t,\vec\sigma')]=i\eta^{\mu\nu}\delta^p(\vec\sigma-\vec\sigma')\,,
    \end{aligned}
\end{equation}
we have
\begin{equation}\label{eq:commutXP}
[A^\mu_{\{m\}},B^\nu_{\{n\}}]=\eta^{\mu\nu}\delta_{\{m+n\}}\equiv\eta^{\mu\nu}\delta_{m_1+n_1,0}\delta_{m_2+n_2,0}\cdots\delta_{m_p+n_p,0}\,.
\end{equation}
The Noether current associated with the global symmetry is
\begin{equation}\label{eq:NoetherCurrent}
    j^\alpha=\frac{\p L}{\p\p_\alpha \bo{X}}\cdot\delta_{\hat{\xi}} \bo{X}-L\xi^\alpha\,.
\end{equation}
By substituting $\xi^\alpha\p_\alpha$ with $-i \hat{l}^i_{\{m\}}$ and $-i \hat{m}_{\{m\}}$, the Noether charge $Q=\int \td^p\sigma j^0$ can be expressed as
\begin{equation}
    \begin{aligned}
        &L^{i,(X)}_{\{m\}}=\sum_{\{n\}}\left(\frac{1+\lambda}{2}m_i-n_i\right):\bo{A}_{\{m-n\}}\cdot\bo{B}_{\{n\}}:\,,\\
        &M^{(X)}_{\{m\}}=-\frac{1}{2}\sum_{\{n\}}\bo{B}_{\{m-n\}}\cdot\bo{B}_{\{n\}}\,.
    \end{aligned}
\end{equation}
Here we introduce the normal-ordering operator $::$, placing all the creation operators to the left of the annihilation operators. The exact meaning of its definition depends on choice of vacuum. And we will address this issue in Sec \ref{sec:QuantumAnomaly}.
From the commutation relations \eqref{eq:commutXP}, we can calculate the algebra generated by these conserved charges. The result, regardless of the quantum anomaly, is
\begin{equation}
    \begin{aligned}
        &[L^{i,(X)}_{\{m\}},L^{j,(X)}_{\{n\}}]=m_j L^{i,(X)}_{\{m+n\}}-n_i L^{j,(X)}_{\{m+n\}}\,,\\
        &[L^{i,(X)}_{\{m\}},M^{(X)}_{\{n\}}]=(\lambda m_i-n_i) M^{(X)}_{\{m+n\}}\,.
    \end{aligned}
\end{equation}

\section{The \texorpdfstring{$b c$}{bc} ghost system}\label{sec:bcghost}

To fully quantize the tensionless brane, we need to consider the partition function in terms of the path integral
\begin{equation}
    Z=\int\tD \hat{V} \tD \bo{X} e^{iS(\hat{V},\bo{X})}=\int\tD \hat{V} \tD \bo{X} e^{iS(\hat{V},\bo{X})}\int\tD\hat{\epsilon}\delta(\hat{V}_{(\hat{\epsilon})}-\hat{V}_{(\hat{0})})\det(\delta_{\hat{\epsilon}} \hat{V}_{(\hat{\epsilon})})\,.
\end{equation}
If there is no anomaly for the Carrollian diffeomorphism, we have $\tD \hat{V}_{(\hat{\epsilon})}\tD \bo{X}_{(\hat{\epsilon})}=\tD \hat{V}\tD \bo{X}$.
As a result,
\begin{equation}
    \begin{aligned}
        Z\rightarrow &Z\Big/\int \tD\hat{\epsilon}\\
        =&\int\tD \hat{V} \tD \bo{X} \delta(\hat{V}-\hat{V}_{(\hat{0})})\det(\delta_{\hat{\epsilon}} \hat{V}) e^{iS(\hat{V},\bo{X})}\\
        =&\int\tD \bo{X} \det(\delta_{\hat{\epsilon}} \hat{V})e^{iS(\hat{V},\bo{X})}|_{\hat{V}=\hat{V}_{(\hat{0})}}\,.
    \end{aligned}
\end{equation}
The determinant $\det(\delta_{\hat{\epsilon}}\hat{V})$ in the path integral can be expressed as the path integral of a $bc$ ghost system
\begin{equation}
    \det(\delta_{\hat{\epsilon}}\hat{V})|_{\hat{V}=\hat{V}_{(\hat{0})}}=\int\tD \hat{b}\tD \hat{c} e^{iS(\hat{b},\hat{c})}\,,
\end{equation}
with
\begin{equation*}
    \begin{aligned}
        S(\hat{b},\hat{c})&=i\left(\frac{1}{2\pi}\right)^p\int \td^d\sigma c^\alpha\left(-\delta^\beta_\alpha V^\gamma\p_\gamma+\p_\alpha  V^\beta+\left(\frac{1}{2}-\frac{\Delta}{d}\right)V^\beta\p_\alpha\right)b_\beta \Bigg|_{\hat{V}=\hat{V}_{(\hat{0})}}\\
        &=i\left(\frac{1}{2\pi}\right)^p\int \td^d\sigma \left(\left(\frac{1}{2}+\frac{\Delta}{d}\right)c^0 \p_0 b_0+c^i\p_0 b_i-\left(\frac{1}{2}-\frac{\Delta}{d}\right)c^i \p_i b_0\right)\,.
    \end{aligned}
\end{equation*}
For convenience, we rescale the component $b_0$ to be $\left(\frac{1}{2}+\frac{\Delta}{d}\right)^{-1}b_0$. Then the action of $bc$ ghost fields becomes
\begin{equation}
    S(\hat{b},\hat{c})=i\left(\frac{1}{2\pi}\right)^p\int \td^d\sigma \left(c^0 \p_0 b_0+c^i\p_0 b_i-\lambda c^i \p_i b_0\right)\,.
\end{equation}
Solving the equations of motion, we have the mode expansions
\begin{equation}
    \begin{aligned}
        &c^0=\lambda \p_i c^i t+\sum_{\{m\}}c^0_{\{m\}}e^{i m_k\sigma^k}\,,\\
        &c^i=\sum_{\{m\}} c^i_{\{m\}}e^{i m_k\sigma^k}\,,\\
        &b_0=\sum_{\{m\}} b_{0,\{m\}}e^{i m_k\sigma^k}\,,\\
        &b_i=\lambda \p_i b_0 t+\sum_{\{m\}}b_{i,\{m\}}e^{i m_k\sigma^k}\,.
    \end{aligned}
\end{equation}
Performing canonical quantization gives us
\begin{equation}\label{eq:CommutCB}
    \begin{aligned}
        &\{c^\alpha(t,\vec\sigma),b_\beta(t,\vec\sigma')\}=(2\pi)^p\delta^\alpha_\beta\delta^p(\vec\sigma-\vec\sigma')\,,\\
        &\{c^\alpha_{\{m\}},b_{\beta,\{n\}}\}=\delta^\alpha_\beta\delta_{\{m+n\}}\,.
    \end{aligned}
\end{equation}

In the rest of this section, we are going to discuss the contribution to $g^{(p)}_\lambda$ generators from the ghost system by using the BRST method. We first determine the BRST charge of the whole system.
The ghost field $c^\alpha$ is in the place of the diffeomorphism under the BRST transformation,
\begin{equation}
    \begin{aligned}
        &\delta_B c^\alpha=i\{c^\alpha,Q_B\}=c^\beta\p_\beta c^\alpha\,,\\
        &\delta_B \bo{X}=-i[\bo{X},Q_B]=c^\alpha\p_\alpha \bo{X}+\frac{\Delta}{d}\p_\alpha c^\alpha \bo{X}\,.
    \end{aligned}
\end{equation}
It is not hard to check that $\delta_B^2 c^\alpha=\delta_B^2 X=0$. To satisfy these BRST transformations, the BRST charge must be of the form
\begin{equation}\label{eq:BRSTcharge}
    Q_B=\int \td^p\sigma\left[\left(\frac{(2\pi)^p}{2}c^0\bo{\Pi}^2+c^i\partial_i \bo{X}\cdot\bo{\Pi}+\frac{1-\lambda}{2}\p_i c^i \bo{X}\cdot\bo{\Pi}\right)-\frac{i}{(2\pi)^p}(c^\alpha\p_\alpha c^\beta b_\beta)\right]\,.
\end{equation}
In terms of the mode expansions, the BRST charge $Q_B$ can be expanded as
\begin{equation}
    \begin{aligned}
        Q_B=&-\sum_{\{m\}}\left(c^0_{\{-m\}}M^{(X)}_{\{m\}}+c^i_{\{-m\}}L^{i,(X)}_{\{m\}}\right)\\
        &+\sum_{\{m\},\{n\}}(m_k-n_k)\Big(\lambda c^0_{\{n\}}c^k_{\{m-n\}}b_{0,\{-m\}}+c^k_{\{n\}}c^0_{\{m-n\}}b_{0,\{-m\}}\\
        &\hspace{3.4cm}+c^k_{\{n\}}c^j_{\{m-n\}}b_{j,\{-m\}}\Big)\,.
    \end{aligned}
\end{equation}
Another way to derive this expression is to write the matter part of the BRST charge first, and the ghost part is determined uniquely by the nilpotency of the BRST charge, regardless of the quantum anomaly, $Q_B^2=0$.
In BRST quantization, the generators of the $\mathfrak{g}^{(p)}_{\lambda}$ algebra can be defined through the anticommutation relations between the antighost modes and the BRST charge,
\begin{equation}
    M_{\{m\}}=-\left\{b_{0,\{m\}},Q_B\right\}\,,\quad L^i_{\{m\}}=-\left\{b_{i,\{m\}},Q_B\right\}\,,
\end{equation}
where $M_{\{m\}}=M^{(X)}_{\{m\}}+M^{(\text{gh})}_{\{m\}}$ and $L^i_{\{m\}}=L^{i,(X)}_{\{m\}}+L^{i,(\text{gh})}_{\{m\}}$ are the total generators. Under this relation, the ghost parts of the generators are given by
\begin{equation}
    \begin{aligned}
        &L^{i,(\text{gh})}_{\{m\}}=-\sum_{\{n\}}n_i :c^\alpha_{\{n\}}b_{\alpha,\{m-n\}}:-\sum_{\{n\}}m_k :c^k_{\{n\}}b_{i,\{m-n\}}:-\lambda\sum_{\{n\}}m_i :c^0_{\{n\}}b_{0,\{m-n\}}:\,,\\
        &M^{(\text{gh})}_{\{m\}}=-\sum_{\{n\}}(\lambda n_k+m_k) c^k_{\{n\}}b_{0,\{m-n\}}\,.\notag
    \end{aligned}
\end{equation}

\section{Quantum anomaly of \texorpdfstring{$\mathfrak{g}^{(p)}_{\lambda}$}{g(p)lambda}}\label{sec:QuantumAnomaly}

In this section, we calculate the quantum anomaly of the algebra $\mathfrak{g}^{(p)}_{\lambda}$ for the matter fields $\bo{X}$ and the $b c$ ghost fields for general $p$ and $\lambda$ with respect to a set of vacua $|\{h\}\>=|h_0,h_1,h_2\>, \{h\}\subset \mathbf{Z}$. The vacuum $|\{h\}\>$ is defined by
\begin{equation}
    \begin{aligned}
        &\bo{A}_{\{m\}}|\{h\}\>=0\quad \mbox{with}~ \sum_i m_i\geq 1-h_0\,,\qquad\bo{B}_{\{m\}}|\{h\}\>=0\quad\mbox{with}~\sum_i m_i\geq h_0\,,\\
        &c^0_{\{m\}}|\{h\}\>=0\quad\mbox{with}~\sum_i m_i\geq 1-h_1\,,\qquad b_{0,\{m\}}|\{h\}\>=0\quad\mbox{with}~\sum_i m_i\geq h_1\,,\\
        &c^i_{\{m\}}|\{h\}\>=0\quad\mbox{with}~\sum_j m_j\geq 1-h_2\,,\qquad b_{i,\{m\}}|\{h\}\>=0\quad\mbox{with}~\sum_j m_{j}\geq h_2\,.\\
    \end{aligned}
\end{equation}
It is natural to choose $h_0=0$ since $\bo{B}_{\{0\}}$ is the momentum operator and $\bo{A}_{\{0\}}$ is the position operator. But we have no reason to set $h_1, h_2$ in advance. Any two different vacua with the same $h_0$ but different $h_1,h_2$ can be related to each other. For example, the vacuum $|\{h'\}\>$ with $h'_1=0$ is related to the vacuum $|\{h\}\>$ with $h_1=1$ by
\begin{equation}\label{eq:translation}
    |\{h'\}\}\>=\prod_{\{m\}\in \text{Set}}b_{0,\{m\}}|\{h\}\>\,,\quad \text{Set}=\left\{\{m\}\Big|\sum_i m_i= 0\right\}\,.
\end{equation}
This relation is realized by the nilpotency of the fermionic-mode operators, i.e. $b_{0,\{m\}}^2=\{b_{0,\{m\}},b_{0,\{m\}}\}/2=0$. In string theory ($p=1$), the right-hand side of \eqref{eq:translation} is composed of a finite number of mode operators acting on the vacuum. Thus, they, $|\{h'\}\}\>$ and $|\{h\}\>$, belong to the same module of a highest weight representation. But when $p>1$, the expression above involves an infinite product of mode operators, and its convergence is not guaranteed, implying that they may belong to different modules of the highest weight representations of $\mathfrak{g}^{(p)}_{\lambda}$.

The quantum anomaly arises from the normal ordering associated with these vacua.
For $p=1$, the $g^{(1)}_\lambda$ algebra
\begin{equation}
    \begin{aligned}
        &[L_{m},L_{n}]=(m-n) L_{m+n}+A(m)\delta_{m+n,0}\,,\\
        &[L_{m},M^{}_{n}]=(\lambda m-n) M^{}_{m+n}\,.
    \end{aligned}
\end{equation}
has a finite anomalous term,
\begin{equation}\label{eq:stringAnomalyX}
    \begin{aligned}
        &A^{(X)}(m)=\frac{D}{6}\left(\frac{3\lambda^2-1}{2}m^2-6h_0(h_0-1)-1\right)m\,,\\
        &A^{(\text{gh})}(m)=-\frac{1}{6}\bigg((6\lambda^2+6\lambda+14)m^2-6h_1(h_1-1)-6h_2(h_2-1)-2\bigg)m\,.
    \end{aligned}
\end{equation}

The algebra $g^{(p)}_\lambda$ for $p>1$ must be of the form
\begin{equation}\label{eq:CCAwithQA}
    \begin{aligned}
        &[L^{i}_{\{m\}},L^{i}_{\{n\}}]=(m_i-n_i) L^{i}_{\{m+n\}}+A^i_1(\{m\})\delta_{\{m+n\},0}\,,\\
        &[L^{i}_{\{m\}},L^{j}_{\{n\}}]=m_j L^{i}_{\{m+n\}}-n_i L^{j}_{\{m+n\}}+A^{i j}_2(\{m\})\delta_{\{m+n\},0}\quad (i\neq j)\,,\\
        &[L^{i}_{\{m\}},M^{}_{\{n\}}]=(\lambda m_i-n_i) M^{}_{\{m+n\}}\,.
    \end{aligned}
\end{equation}
Thus, we may calculate the anomalous terms $A^i_1(\{m\}),A^{i j}_2(\{m\})$ by
\begin{align*}
    \left[L^i_{\{m\}},L^i_{\{-m\}}\right]|\{h\}\>&=&A^i_1(\{m\})|\{h\}\>\\
    \left[L^i_{\{m\}},L^j_{\{-m\}}\right]|\{h\}\>&=&A^{i j}_2(\{m\})|\{h\}\> ~~\mbox{for $j\neq i$.}
\end{align*}
The anomalous terms involve some infinite sums. For example, the anomaly from the fields $X$ is
\begin{equation}\label{eq:anomalyX}
        A_1^{i,(X)}(\{m\})=\sum_{\{n\}\in\text{Set}(h_0)}\left(\frac{1+\lambda}{2}m_i-m_i-n_i\right)\left(\frac{1+\lambda}{2}m_i+n_i\right)\,,
\end{equation}
with
\begin{equation}
    \text{Set}(h_0)=\left\{\{n\}\Big|-\sum_i m_i+h_0\leq\sum_i n_i\leq-1+h_0\right\}
\end{equation}
Denote $\sum_i n_i=N$, the range of the sum can be written as a finite sum times some infinite sum,
\begin{equation}\label{eq:sumDecomp}
    \sum_{\text{Set}(h_0)}=\sum_{N=-\sum_i m_i+h_0}^{-1+h_0}\sum_{n_1=-\infty}^\infty\cdots\sum_{n_{p-1}=-\infty}^\infty\Big|_{n_p=N-\sum_{i=1}^{p-1}n_i}
\end{equation}
For example, assuming $i\neq p$ without loss of generality, the anomalous terms \eqref{eq:anomalyX} can be written as
\begin{equation}
    \begin{aligned}
        A_1^{i,(X)}(\{m\})&=\sum_{n_i=-\infty}^\infty\left(\frac{1+\lambda}{2}m_i-m_i-n_i\right)\left(\frac{1+\lambda}{2}m_i+n_i\right)\sum_{N=-\sum_i m_i+h_0}^{-1+h_0}\prod_{j\neq i}^{p-1}\sum_{n_j=-\infty}^\infty 1\\
        &=\left(\sum_i m_i\right)\left(\sum_{n=-\infty}^\infty 1\right)^{p-2} \sum_{n_i=-\infty}^\infty\left(\frac{1+\lambda}{2}m_i-m_i-n_i\right)\left(\frac{1+\lambda}{2}m_i+n_i\right)\,.
    \end{aligned}
\end{equation}
Note that \eqref{eq:sumDecomp} only holds for $p>1$. If $p=1$, it is a finite sum, and we have
\begin{equation}
    A^{(X)}(m)=\sum_{n=-m+h_0}^{-1+h_0} \left(\frac{1+\lambda}{2}m-m-n\right)\left(\frac{1+\lambda}{2}m+n\right)\,,
\end{equation}
which exactly gives the first result of Eq.\eqref{eq:stringAnomalyX}.

Thus, when $p>1$, all the anomalous terms can finally be translated into something including at least one factor of the form,
\begin{equation}
    S(l)=\sum_{n=-\infty}^\infty n^l .\qquad  \mbox{($l$ is some non-negative integer)}
\end{equation}
Exploiting the Riemann Zeta function $\xi(s)\equiv\sum_{n=1}^\infty1/n^s$, we can regularize the
expression above. If $l$ is an odd positive integer, $S(l)=0$ by parity. And if $l$ is an even non-negative integer, $S(l)=\delta_{l,0}+2\xi(-l)=\delta_{l,0}+2\delta_{l,0}\xi(0)=0$. Anyway, the divergent anomalous terms in the form of $S(l)$ are zero after the Riemann Zeta function regularization.

To make this point more explicit, we illustrate with two representative cases.
For $p>2$, the factorized expression for $A_1^{i,(X)}$ contains an overall factor $(\sum_{n=-\infty}^\infty 1)^{\,p-2}=[S(0)]^{p-2}$, which is already zero under the zeta regularization, so the entire anomalous term vanishes irrespective of the remaining factor.
For $p=2$, there is no such divergent overall factor; however, the single sum $\sum_{n_i=-\infty}^\infty(\cdots)$ is a polynomial in $n_i$. Expanding this polynomial yields a linear combination of $S(l)$ with $l\ge 0$, and each $S(l)$ vanishes under the same regularization.
The same reasoning extends to the cross-terms arising from the multiple sums in the decomposition~\eqref{eq:sumDecomp}: every term can be written as a product of factors of the form $S(l)$, and no finite residual contribution survives the zeta regularization.

This vanishing of the anomalous terms in $p>1$ case is in accordance with the fact that the Gelfand-Fuchs cohomology $H^2(\text{Vect}(T^p),\mathbb{C})$ is trivial \cite{fuchs1986cohomology}, and there is no scalar central extension of the algebra involving $L^i_{\{m\}}$. And we also expect this for the algebra $\mathfrak{g}^{(p)}_{\lambda}\cong \text{Vect}(T^p)\ltimes_\lambda C^\infty(T^p)$ by the proposition 1 in Section 3 of the papers \cite{GAO2016107, Buzaglo:2024jqo}. A generalized Abelian extension is given by \cite{BermanMoody1994}, but that is beyond the scope of our interest.

\section{Conclusions and discussions}\label{sec:conclusions}

For general $p$ and the chosen vacua, we obtain the quantum anomaly with respect to the algebra $\mathfrak{g}^{(p)}_{\lambda}$. The consistency of a quantum gauge system then requires the total central charge to vanish. When $p=1$, the leading term gives the constraint
\begin{equation}
    \frac{3\lambda^2-1}{2}D-[6(2+\lambda+\lambda^2)+2]=0\,,
\end{equation}
which has two nontrivial and sensible solutions,
\begin{equation}
    D=14 \quad\text{when}\quad \lambda=-1\,,
\end{equation}
and
\begin{equation}
    D=26 \quad\text{when}\quad \lambda=1\,.
\end{equation}
The lower-order term in the quantum anomaly can be absorbed into the normal order constant (or equivalently redefine $L_0\rightarrow L_0-a_L$).

For the $p>1$ cases, the quantum anomaly vanishes automatically after the Riemann Zeta function regularization. Thus, there is no constraint on the critical dimension for the brane theories.

We emphasize that this result only implies that the quantum consistency of the $\mathfrak{g}^{(p)}_\lambda$ algebra does not restrict the spacetime dimension; other potential quantum instabilities of the brane theory, such as negative-norm states, global anomalies, or dynamical instabilities, still need to be examined separately.

In string theory, the origin of the parameter $\lambda$ traces back to different Ultra-relativistic(UR) limits. The action for the $bc$ ghost returns to the inhomogeneous case of \cite{Chen:2023esw} when $\lambda=1$ and the homogeneous case when $\lambda=0$. In \cite{Chen:2023esw}, the authors claimed that the homogeneous $bc$ ghost does not admit a BRST charge and thereafter does not make sense. This claim is not correct as they did not find the correct symmetry algebra $g^{(p)}_{\lambda=0}$ there. As shown in this work, the homogeneous $bc$ ghost corresponds to the Carrollian conformal symmetry with $\lambda=0$ rather than with $\lambda=1$, the symmetry generators $L_m, M_m$ should be modified for both the matter sector and the ghost sector, and the correct BRST charge is constructed as \eqref{eq:BRSTcharge} in this manuscript. And it is not hard to check that the result in \cite{Chen:2023esw} studied for the bosonic tensionless string is exactly the special case $p=1,\lambda=1$ in this work.

A subtle point concerns the critical dimension $D=14$ obtained at $\lambda=-1$. Recall that $\lambda$ is not a free parameter but follows from the conformal dimension $\Delta$ through $\lambda=(\frac12-\frac{\Delta}{d})/(\frac12+\frac{\Delta}{d})$; solving $\lambda=-1$ requires $|\Delta|/d\to\infty$, a limit inaccessible for any finite $\Delta$. Nevertheless, the quantities entering the quantum theory all remain finite in this limit thanks to a systematic $0\cdot\infty$ cancellation. From the gauge-preserving condition one finds $\epsilon^0=\lambda\partial_i f^i t+g$, giving $\partial_\alpha\epsilon^\alpha=(1+\lambda)\partial_i f^i$, which vanishes as $\lambda\to-1$ and cancels the divergent factor $\Delta/d$ in the combined gauge transformation of $X$. The ghost field inherits the same structure $c^0=\lambda\partial_i c^i t+\cdots$, so $\partial_\alpha c^\alpha=(1+\lambda)\partial_i c^i\to0$. The BRST charge~\eqref{eq:BRSTcharge} contains the coefficient $\frac{1-\lambda}{2}$ multiplying $\partial_i c^i X\!\cdot\!\Pi$; this coefficient is not singular because of the identity $\frac{1-\lambda}{2}=\frac{\Delta}{d}(1+\lambda)$, which tends to $1$ as $|\Delta|/d\to\infty$. Consequently the BRST charge, the algebra $\mathfrak{g}^{(1)}_{-1}$, and the anomaly calculation in Sec.~\ref{sec:QuantumAnomaly} are all perfectly regular when evaluated at $\lambda=-1$, and $D=14$ follows as a mathematically consistent critical dimension. The obstruction is not algebraic but lies in the Faddeev-Popov derivation of the ghost action: before the rescaling $b_0\to(\frac12+\frac{\Delta}{d})^{-1}b_0$, the ghost kinetic coefficients $\frac12\pm\frac{\Delta}{d}$ diverge, and the rescaling itself becomes an infinite field redefinition whose Jacobian in the path integral measure is not controlled. Whether $\lambda=-1$ can be physically realized---for instance through a different gauge-fixing prescription or by treating $\lambda$ as an intrinsic parameter of the symmetry algebra---remains an open question.

We note that the algebra $\mathfrak{g}^{(1)}_\lambda$ can be understood as a linear combination of the standard BMS$_3$ algebra and the Carroll-Weyl symmetry generators; a detailed analysis of this connection will be presented in a forthcoming work.

If we consider the tensionless super-brane, we expect some SUSY $\mathfrak{g}^{(p)}_{\lambda}$ algebra as in the string case \cite{Bagchi:2022owq, Bagchi:2025jgu}. We hope to address that in the future. And since we have obtained the BRST charge for the tensionless brane, discussing its spectrum through BRST quantization \cite{Figueroa-OFarrill:2025njv} is also a very interesting question.

In the string case $(p=1)$, the action of Carrollian string contains magnetic sectors of Carrollian conformal scalar theory \cite{Chen:2024voz}, which are T-dual to the electric sectors in the action of tensionless string. From this point of view, the $\mathfrak{g}^{(1)}_{\lambda}$ algebra also appears in the magnetic sector when $p=1$. For the higher-dimensional case ($p>1$), the action of the tensionless brane is also composed of the electric sectors of Carrollian conformal scalar theory, but there is no evidence supporting the T-duality so far. Then, whether the magnetic sector or other models exhibit $\mathfrak{g}^{(p)}_{\lambda}$ algebra is another interesting question.

\acknowledgments

We thank Yu-fan Zheng for meaningful discussions. In particular, he provides insight from point-particle models. The work is partly supported by NSFC Grant No. 12275004 and No. 12588101.

\appendix

\bibliographystyle{iopart-num}
\bibliography{biblio.bib}

\end{document}